\documentclass[showpacs, showkeys, aps,prl,preprint,groupedaddress]{revtex4-1}
\usepackage{graphicx}
\usepackage{mathrsfs}
\usepackage{amssymb,amsmath}
\preprint{USTC-ICTS-12-08}

\begin{document}

\title{$\omega  =-1$  crossing  in  quintessence  models  in  Lyra's  geometry}
\author{Hoavo      Hova\email{hovhoav@mail.ustc.edu.cn}      and      Huanxiong
Yang\email{hyang@ustc.edu.cn}}

\affiliation{Interdisciplinary center for theoretical study, University of
Science and Technology of China, Hefei, 230026, P. R. China}

\date{\today}

\begin{abstract} We study  the cosmology of quintessence models  in an extended
theory of  gravity in Lyra's  geometry. By analyzing the  possible interactions
between the quintessence scalar and  the intrinsic displacement field in Lyra's
geometry,  we  obtain the  closed  form  solutions  of the  modified  Friedmann
equations for four  classes of quintessence models. Though the  presence of the
geometrical  displacement  field promises  the  possibility  for the  effective
equation of  state $\omega$  of the quintessence-displacement  mixture crossing
the  cosmological constant  boundary,  the reliable  quintessence scenarios  in
Lyra's geometry  with stable perturbation  modes are  still those in  which $-1
\leq \omega \leq 1$.
\end{abstract}

\maketitle

\section{Introduction}

Recent   cosmic   observations    \cite{Riess,   Perlmutter1999,   Tegmark2004,
Tegmark2004a,  Tegmark2006,   Seljak2005,  AdelmanMcCarthy2006,  Abazajian2003,
Abazajian2004, Abazajian2005, Spergel2007,  Page2007, Hinshaw2007, Jarosik2007}
have indicated that our Universe is  undergoing an accelerated expansion at the
present epoch.   The cause for such  a cosmic acceleration is  attributed to an
unknown  dominant energy  component, dubbed  \emph{dark energy},  with negative
pressure  generating thus  repulsive gravitational  forces that  counteract the
attractive  forces   produced  by   radiation,  baryons   and  the   cold  dark
matter. However,  the exact nature  of dark  energy is currently  a significant
part of the realm of speculations. Some  believe that dark energy is the energy
of  the quantum  vacuum, modelled  by  the cosmological  constant $\Lambda$  of
general relativity. Interpreting  dark energy as a  cosmological constant means
that  the  density of  dark  energy  is  uniform  throughout the  universe  and
invariable in time. This is the simplest explanation for dark energy, which was
introduced by Einstein for  building a static universe but has  a good fit with
the available  data of  the current cosmological  observations. If  dark energy
takes this form, it is a fundamental property of the universe. The cosmological
constant thus faces two fundamental problems in physics, namely the fine-tuning
and coincidence puzzles.

The late-time cosmic acceleration may alternatively be driven by a dynamic dark
energy  which  could  be  a   time  evolving  and  spatially  dependent  scalar
field. Lots  of such dynamic dark  energy models have been  proposed, which are
roughly  classified   into  three  categories:   quintessence  \cite{Ratra1988,
Peebles2003,  Caldwell1998, Wetterich1988,  Coble1997, Turner1997,  Boyle2002},
phantom \cite{Onemli2002,  Caldwell2002, Carroll2003,  Onemli2004, Brunier2005,
Kahya2007,   Kahya2010}  and   quintom  \cite{Feng2005,   Guo2005,  Vikman2005,
Zhao2005, Wei2005, Capozziello2006, Cai2007, Cai2010a}. In quintessence models,
a scalar field $\varphi$ with a canonical kinetic energy and a self-interaction
potential energy $V(\varphi)$  is supposed to be minimally  coupled to Einstein
gravity. In a flat Robertson-Walker  background the quintessence scalar behaves
as a perfect  fluid with an evolving equation-of-state  (EoS) parameter $\omega
=p/\rho$ lying  in the range  $-1 \leq \omega \leq  1$.  In phantom  models the
quintessence  is replaced  by a  ghost scalar  of which  the kinetic  energy is
negative   and  $\omega   <-1$.   Due   to  the   no-go  theorem   proposed  in
Ref.\cite{Vikman2005, Hu2005,  Caldwell2005, Zhao2005, Kunz2006,  Xia2008}, the
model  buildings  of  quintom  dark   energy  are  generally  very  complicated
\cite{Cai2010a}. The simplest  quintom model is composed of  two scalar fields,
one is  a quintessence scalar  and another a phantom  \cite{Feng2005, Guo2005}.
Quintom models characterize  themselves by the property that  the effective EoS
parameter  can cross  the cosmological  constant boundary  $\omega =-1$,  which
makes them to fit the observational data better \cite{Cai2010a}.

Crossing the $\omega=-1$  divide in a dynamic dark energy  model is bewitching.
However,  the emergence  of  a phantom  mode with  negative  kinetic energy  in
quintom  models brings  about  great embarrassment  in  understanding it.   The
consistence  coming from  the Null  Energy  Condition in  physics requires  the
kinetic  energy of  a normal  scalar field  not to  be negative,  otherwise the
theory might be  unstable and unbounded.  Therefore, it is  worthwhile to study
the mechanism of  removing the phantom field from the  quintom models. In fact,
there has lots of such attempts to investigate the possibility of $\omega = -1$
crossing in  quintessence like models  \cite{Cai2007,Nicolis2009, Deffayet2009,
Deffayet2010}. It has been empirically realized  that to cross the $\omega =-1$
barrier and remove  ghost mode at the  same time, the model  building should be
involved in  either modifying  the general theory  of Einstein's  relativity or
introducing some higher derivative terms for the scalar fields. For example, In
the so-called Galileon cosmology \cite{Nicolis2009, Deffayet2009, Deffayet2010}
of a scalar field, the higher  derivatives of operators are introduced into the
Lagrangian  but the  equation of  motion of  the scalar  remains of  the second
order.  The Galileon models can have  $\omega =-1$ crossing without ghost modes
involved. It goes without saying, however, that these models are generally very
complicated to deal with.

Of the  modification attempts beyond  Einstein's general theory  of relativity,
there is  an extended  theory of  gravity (ETG) based  on the  so-called Lyra's
geometry \cite{Lyra, Scheibe, Beesham, Sen1957, Halford1970}. As is well known,
the general  theory of  relativity is  a theory of  gravity built  on (pseudo-)
Riemannian geometry. Lyra's  geometry is a modification  of Riemannian geometry
where a \emph{gauge function} is introduced.  Due to the presence of this gauge
function  on  the structure-less  manifold,  an  extra geometrical  ingredient,
\emph{i.e.}, the displacement vector $\beta_\mu$  arises in Lyra's geometry. It
is remarkable that the connection in  both Riemannian and Lyra's geometries are
metric preserving.  The  extended theory of gravity in Lyra's  geometry is much
motivated by  the fact  that it  could predict the  same effects  as Einstein's
general relativity  within observations limits,  as far as the  classical Solar
System, as well  as tests based on  the linearised form of  the field equations
\cite{Halford1970,   Pradhan2004,    Soleng1987,   Soleng1988,   Matyjasek1993,
Mohanty2006, Mohanty2007, Mohanty2007a,  Mohanty2009, Rahaman2002, Rahaman2003,
Rahaman2003a}. The extended theory of  gravity on Lyra's geometry distinguishes
itself by  the fact  that it is  a scalar-tensor theory  of gravity,  where the
scalar  field  is not  alien,  but  intrinsic  to the  geometry  \cite{Scheibe,
Sen1957}.   Moreover,  in  the  so-called  \emph{normal  gauge}  \cite{Sen1960,
Sen1971, Sen1972,  Manoukian1972, Hudgin1973},  a constant  displacement vector
can play the role of a positive cosmological constant (in the presence of other
cosmic matter ingredients)  \cite{Halford1972, Shchigolev2012, Hova2012}, which
is in  contrast to general relativity  where the cosmological constant  must be
added in an \emph{ad hoc} manner into the gravitational field equations.

In  this  paper,  we  study   the  $\omega=-1$  crossing  possibility  in  some
quintessence models in the framework of the ETG in Lyra's geometry. Despite the
impossibility for  a canonical quintessence  model to cross the  phantom divide
$\omega   =-1$   in   Einstein's    gravity   in   pseudo-Riemannian   geometry
\cite{Vikman2005, Hu2005, Zhao2005},  the existence of a  displacement field in
the ETG in  Lyra's geometry does probably modify the  effective distribution of
the cosmic  fluids so  that the  EoS parameter of  the quintessence  scalar may
cross this  boundary.  Aimed  at finding  the exact  solutions of  the modified
Fridemann equations in  a flat Robertson-Walker background,  we propose several
candidate  interactions between  the quintessence  scalar and  the displacement
field which  have simple mathematical  expressions. For some of  these possible
interactions, crossing $\omega  =-1$ barrier for quintessence  models in Lyra's
geometry is  available. Unfortunately,  crossing this  phantom divide  in these
models  will, without  any  exception,  give rise  to  the  instability of  the
relevant perturbations. The reliable  quintessence scenarios in Lyra's geometry
are still those in which $-1 \leq \omega \leq 1$.

The paper is organized as follows.  Section II begins with a brief introduction
to modified Einstein equations in the considered ETG and its application to the
cosmology  of  a  quintessence  model in  a  flat  Robertson-Walker  background
spacetime. By analyzing  the equation of motion of the  quintessence scalar, we
determine phenomenologically several candidate interactions between this scalar
and the displacement field. In Section  III we study the quintessence cosmology
for each of the candidate  interaction terms. The self-interaction potential of
the  quintessence scalar  is not  given a  prior, which  is defined  during the
process solving the modified Friedmann  equations, motivated by the requirement
to have closed  form solutions to these equations. Among  the four quintessence
models  proposed,  three of  them  naively  allow  $\omega =-1$  crossing.   By
requiring the squared  sound speed of the quintessence  scalar preserves finite
and non-negative  during its evolution,  the possibility for  $\omega$ crossing
the phantom  divide is excluded.  It  turns out that the  reliable quintessence
models in Lyra's geometry are also characterized by inequalities $-1 \leq \omega
\leq 1$, similar  to those in pseudo-Riemannian geometry.  We  conclude in Section
IV with a  summary of the results  and some remarks. For simplicity  we work in
the Planck units $c =\hbar = \kappa^{2}=1$ throughout the paper.

\section{Quintessence and Accelerated Expansion}

The  quintessence models  in  an extended  theory of  gravity  (ETG) in  Lyra's
geometry, in the so-called normal gauge, is described by the following modified
Einstein gravitational field equations \cite{Sen1957}:
\begin{equation}
\label{eq:1}
G_{\mu\nu}=T^{\varphi}_{\mu\nu}+{\mathscr T}_{\mu\nu},
\end{equation}
where,
\begin{equation}
\label{eq:2}
{\mathscr T}_{\mu\nu}=-\frac{3}{2}\left(\beta_{\mu}\beta_{\nu} -\dfrac{1}{2}
  g_{\mu\nu} \beta_{\lambda} \beta^{\lambda} \right),
\end{equation} is an intrinsic geometrical  stress tensor, corresponding to the
existence  of the  displacement  vector $\beta_{\mu}$  which  emerges from  the
integrability  condition  of  length  of  a  vector  under  parallel  transport
\cite{Lyra,  Sen1957}. $T^{\varphi}_{  \mu\nu}$  is the  stress  tensor of  the
quintessence  scalar $\varphi$  which  is assumed  to  have a  self-interaction
potential $V(\varphi)$ and be canonically coupled to gravity,
\begin{equation}
\label{eq:3}
T^\varphi_{\mu\nu} = \nabla_{\mu}\varphi \nabla_{\nu}\varphi -\frac{1}{2}g_{\mu\nu}\nabla_{\lambda}\varphi \nabla^{\lambda}\varphi -g_{\mu\nu}V(\varphi )
\end{equation}  We further  assume  that, at  present  epoch, the  quintessence
scalar $\varphi$ dominates  over other cosmic fluids such as  baryonic dust and
radiation. The  displacement vector  $\beta_\mu$ is allowed  to be  a time-like
4-vector field \cite{Sen1957, Sen1960, Sen1971, Sen1972},
\begin{equation}
\label{eq:4}
\beta_{\mu}=\left(\beta(t),\,0,\,0,\,0 \right)
\end{equation}
Its unique non-vanishing component $\beta(t)$ can either be a constant or time-dependent.

In a  flat \emph{Friedmann-Lema\^{i}tre-Robertson-Walker} background  $ds^{2} =
-dt^2  +  a^{2}  d\vec{x}^{2}$,  the   modified  Einstein  equations  given  in
Eq.(\ref{eq:1}) become:
\begin{eqnarray}
\label{eq:5}       3H^{2}        =\frac{1}{2}\dot{\varphi       }^{2}+V(\varphi
)-\frac{3}{4}\beta^{2}(t)\\
\label{eq:6} 2\dot{H}+3H^{2}  = -\frac{1}{2}\dot{\varphi }^{2} +  V(\varphi ) +
\frac{3}{4}\beta^{2}(t)
\end{eqnarray} where an  overdot denotes derivative with respect  to the cosmic
time $t$.  The equation of motion  of the quintessence scalar, which comes from
the Bianchi  identities of modified  Einstein equations, \emph{i.e.},  from the
compatibility of Eq.(\ref{eq:5}) with Eq.(\ref{eq:6}), reads,
\begin{equation}
\label{eq:7}   \dot{\varphi}\left(   \ddot{\varphi   }   +3H   \dot{\varphi   }
+V_{,\varphi }\right) = \frac{3}{4} \left(\dot{\theta} + 6H\theta \right),
\end{equation}   In   Eq.(\ref{eq:7})  $V_{,\varphi}=\frac{dV}{d\varphi}$   and
$\theta \equiv  \beta^{2}$ ($\theta$  is also referred  to as  the displacement
field).   Eq.(\ref{eq:7})   implies  that   the  displacement  field   and  the
quintessence scalar $\varphi$ interact as the universe evolves.

The mixture  of the  quintessence scalar $\varphi$  and the  displacement field
$\theta$ is conventionally viewed as a  perfect fluid, whose energy density and
pressure are defined by,
\begin{eqnarray}
\label{eq:8}
&   & \rho =\frac{1}{2}\dot{\varphi }^{2} + V(\varphi )-\frac{3}{4}\theta(t)\\
\label{eq:9}
&   & p = \frac{1}{2}\dot{\varphi }^{2} - V(\varphi )-\frac{3}{4}\theta(t)
\end{eqnarray}
where $\rho_{\varphi} = \frac{1}{2}\dot{\varphi }^{2} + V(\varphi )$ and
$p_{\varphi} = \frac{1}{2}\dot{\varphi }^{2} - V(\varphi )$ are respectively
the energy density and pressure of the quintessence scalar $\varphi$. The  real  displacement vector  $\beta_\mu$ (or  positive
$\theta$)  that  is a  necessary  geometrical  ingredient in  Lyra's  geometry,
nevertheless, behaves as  an exotic cosmic matter with  negative energy density
and negative pressure, $\rho_{\theta} = p_{\theta} = -\frac{3}{4}\theta$. In the absence of quintessence scalar, the effective EoS
parameter  of the  displacement field  is  equal to  $\omega_\theta =1$,  which
corresponds  to  the  so-called  stiff fluid  \cite{Shchigolev2012}.  When  the
quintessence scalar exists, the displacement field would probably play the role
of a phantom \cite{Onemli2002, Caldwell2002,  Carroll2003}, so that the mixture
probably behaves as an effective quintom \cite{Feng2005, Guo2005} to cause the late time
accelerated expansion of our universe.

The effective EoS parameter of the mixed fluid is,
\begin{equation}
\label{eq:10}
\omega := \frac{p }{\rho} \,
=\frac{2\dot{\varphi }^{2}-4V(\varphi )-3\theta(t)}{2\dot{\varphi }^{2}
+ 4V(\varphi ) -3\theta(t)}
\end{equation}  If there  were no  displacement field  $\theta$ (as  in general
relativity based on pseudo-Riemannian geometry),  the EoS parameter $\omega$ would
only include  the contribution of  quintessence scalar $\varphi$, and  $-1 \leq
\omega  \leq  1$.  This  is  not  the  case  in  the  ETG  in  Lyra's  geometry
\cite{Shchigolev2012,  Hova2012}.   In  a  Lyra  manifold   with  the  positive
displacement field  $\theta$, the universe  evolves between $\omega  =1$ (stiff
matter), where either  the kinetic term dominates or both  kinetic term and the
displacement   field  dominate,   and   a   phantom  regime   \cite{Onemli2002,
Caldwell2002, Carroll2003},  where $\omega \le  -1$, provided that  the kinetic
term  of  the scalar  field  is  negligible.  The interesting  possibility  for
$\omega$  crossing  the  cosmological  constant boundary  seems  plausible.  To
clarify such a possibility, the  mechanism describing the possible interactions
between the quintessence  scalar and the displacement field  is required. Owing
to the  lack of  such a  mechanism, in  this paper,  we determine  the relevant
interaction terms by a simple dimensional analysis. The interaction strength is
inevitably  encoded into  a  few phenomenological  parameters. Following  other
works  \cite{Limiao2009,  Amendola2010},  we interpret  Eq.(\ref{eq:7})  as  an
effective energy conservation equation, and recast it as:
\begin{eqnarray}
\label{eq:11}
\dot{\varphi }\left( \ddot{\varphi }+3H\dot{\varphi }+V_{,\varphi}\right)={\mathscr Q},\\
\label{eq:12}
\frac{3}{4}\left(\dot{\theta}+6H\theta \right)={\mathscr Q},
\end{eqnarray} where  ${\mathscr Q}$  stands for  the required  interaction. By
dimensional  analysis,  ${\mathscr Q}$  is  generally  of the  form  ${\mathscr
Q}=\alpha_{1}  \rho_{1}+\alpha_{2}  \rho_{2}  + \cdots$,  with  $\rho_{i}$  the
different  energy components.  The  coupling parameters  $\alpha_{i}$ have  the
dimension  of  Hubble parameter  $H$  or  $\dot{\varphi}$,  and should  not  be
simultaneously  set to  zero. In  the models  under consideration  the possible
candidates of  $\rho_{i}$ are  $\dot{\varphi}^{2}$, $V(\varphi)$  and $\theta$.
We will simply assume that,
\begin{equation}
\label{eq:13}       {\mathscr      Q}       =3c      H\dot{\varphi}^{2}       +
\frac{3b}{2\sqrt{6}}\dot{\varphi} \theta  + \frac{9}{2}\tilde{b}H\theta +  6f H
V(\varphi)\,
\end{equation}  where $c$,  $b$,  $\tilde{b}$ and  $f$  are some  dimensionless
coupling constants.  The choice for  the interaction terms  in Eq.(\ref{eq:13})
allows  us, not  only  to  explain the  present  accelerated  expansion of  our
universe, but to have closed form solutions of the modified Friedmann equations
also.

\vspace{3mm}

We now  proceed to study  the cosmological  consequences emerging from  each of
these interactions. Instead of $t$, we  will use the e-folding number $x=\ln a$
as the time variable for convenience,  with $x=0$ representing the present time
($  a(0)  =1$).  The  modified  Friedmann  equations  in  Eqs.(\ref{eq:5})  and
(\ref{eq:6}) are recast as:
\begin{eqnarray}
\label{eq:14} \frac{d H^2}{d x}+6H^{2}= 2 V \\
\label{eq:15}  \theta =  \frac{2}{3} H^2  \bigg(\frac{d\varphi}{dx} \bigg)^2  +
\frac{4}{3} V -4H^2
\end{eqnarray}  Similarly,  Eqs.(\ref{eq:11})  and (\ref{eq:12})  that  can  be
viewed as the equations of motion of quintessence scalar and displacement field
become:
\begin{eqnarray}
\label{eq:16} & &\frac{d\varphi}{dx}\bigg[ H^2 \frac{d^2\varphi}{dx^2} + \bigg(
3H^2    +   \frac{1}{2}    \frac{dH^2}{dx}    \bigg)   \frac{d\varphi}{dx}    +
V_{,\varphi}\bigg]=\tilde{\mathscr Q} \\
\label{eq:17}  &  &\frac{3}{4}\bigg(  \frac{d\theta}{dx}  +  6\theta  \bigg)  =
\tilde{\mathscr Q}
\end{eqnarray} where,
\begin{equation}
\label{eq:18} \tilde{\mathscr  Q} =3c H^2 \bigg(  \frac{d\varphi}{dx}\bigg)^2 +
\frac{3b}{2\sqrt{6}}    \bigg(    \frac{d\varphi}{dx}     \bigg)    \theta    +
\frac{9}{2}\tilde{b} \theta + 6 f V(\varphi)\,
\end{equation} Relying on the fact that Eq.(\ref{eq:7}) is the Bianchi identity
of  the  modified  Friedmann   equations  (\ref{eq:5})  and  (\ref{eq:6}),  any
non-degenerate combination of three  of equations (\ref{eq:14}), (\ref{eq:15}),
(\ref{eq:16}) and (\ref{eq:17}) will be mathematically equivalent.

\vspace{5mm}
\begin{description}
\item[Case\,1 \,$c\ne 0$ but $b= \tilde{b} = f= 0$\, ]
\end{description} In this case, Eq.(\ref{eq:16})  pretends to decouple from the
displacement  field $\theta$and  becomes  an effective  equation  of motion  of
quintessence scalar $\varphi$,
\begin{equation}
\label{eq:19}  H^2 \frac{d^2\varphi}{dx^2}  + \bigg[  3(1-c) H^2  + \frac{1}{2}
\frac{dH^2}{dx} \bigg] \frac{d\varphi}{dx} + V_{,\varphi} = 0
\end{equation} Equivalently,
\begin{equation}
\label{eq:20} H^2 \bigg(  \frac{d^2\varphi}{dx^2} -3c \frac{d\varphi}{dx}\bigg)
+ \bigg( V_{,\varphi} + V \,\frac{d\varphi}{dx}\bigg) = 0
\end{equation}  when Eq.(\ref{eq:14})  is taken  into account.  In view  of the
mathematical structure of Eq.(\ref{eq:20}), we assume that the self-interaction
quintessence potential $V(\varphi)$ is defined by,
\begin{equation}
\label{eq:21} V_{,\varphi} + V \,\frac{d\varphi}{dx} = 0
\end{equation} in the models under consideration. The evolution of quintessence
scalar in terms of efolding time $x$  turns out to be independent of the Hubble
parameter,
\begin{equation}
  \label{eq:22} \frac{d^2\varphi}{dx^2} -3c \frac{d\varphi}{dx} = 0
\end{equation}  By  assigning  the  initial conditions  $\varphi(0)  =  p$  and
$d\varphi(0)/{dx} = q$,  with $p,\, q$ two outstanding constants,  we can solve
Eq.(\ref{eq:22}) analytically. The solution reads,
\begin{equation}
\label{eq:23} \varphi(x) = p + q(e^{3c x} -1)
\end{equation}  The constant  $p$  does  not matter  whether  the scalar  field
$\varphi$  is a  quintessence or  a phantom,  because it  is irrelevant  to the
kinetic  energy  of  the field.  So,  we  set  $p=0$  for simplicity  from  now
on. Substitution of Eq.(\ref{eq:23}) into Eq.(\ref{eq:21}) leads to:
\begin{equation}
\label{eq:24}  V(x)  =  r   H_0^2  \exp\left[-\frac{3}{2}  c  q^2  \big(e^{6cx}
-1\big)\right] \,
\end{equation} In  Eq.(\ref{eq:24}) we  have used a  constant $r$  labeling the
present value  $r H_0^2$ of the  quintessence potential, with $H_0$  the Hubble
parameter at  present epoch. The  potential can be  expressed as a  closed form
function of the quintessence scalar $\varphi$ itself,
\begin{equation}
  \label{eq:25}     V(\varphi)     =    r     H_0^2     \exp\left[-\frac{3}{2}c
\varphi\big(\varphi + 2q\big) \right]
\end{equation} If $c<0$ and $q=0$, such  a $V(\varphi)$ could be interpreted as
the tachyon potential describing the excitation of massive scalar fields on the
anti-D branes \cite{Garousi2004, Garousi2005}, with $r H_0^2$ the brane tension and $-3 c $ the
mass squared of  the field $\varphi$ in the Planckian  units. However, $q=0$ is
forbidden in our case. The  physics behind the proposed potential (\ref{eq:25})
is an open issue.

With Eq.(\ref{eq:25}), we  can obtain the evolution of Hubble  parameter $H$ by
solving Eq.(\ref{eq:14}). The result is,
\begin{equation}
\label{eq:26}      H^2(x)       =      H_0^2      e^{-6x}       \left\{1      +
\frac{r}{3c}\sqrt[c]{\frac{2}{3c      q^2}}\,      e^{3c     q^2/2}      \bigg[
\Gamma\bigg(\frac{1}{c},  \frac{3c   q^2}{2}\bigg)  -  \Gamma\bigg(\frac{1}{c},
\frac{3c q^2}{2} e^{6 c x}\bigg) \bigg]\right\}
\end{equation}  where $\Gamma(\tau,  z)$  is the  \emph{upper incomplete  Gamma
function} \cite{Arfken2005},
\begin{equation}
  \label{eq:27} \Gamma(\tau, z)  = \int\limits_z^{+\infty} d\zeta \,\zeta^{\tau
-1} e^{-\zeta}
\end{equation} Furthermore,  employment of Eqs.(\ref{eq:25})  and (\ref{eq:26})
in Eq.(\ref{eq:15}) yields,
\begin{eqnarray}
\label{eq:28}  & \theta(x)  & \,  = \,\frac{4}{3}r  H_0^2 \exp\bigg[\frac{3}{2}
cq^2 (1- e^{6 c x})\bigg] + 2 H_0^2  (3c^2 q^2 e^{6 c x} -2) e^{-6 x} \nonumber
\\  &  & \,\,\,\,\,\,\,\,+  \,\frac{2  r}{3c}H_0^2  (3c^2  q^2  e^{6 c  x}  -2)
\sqrt[c]{\frac{2}{3cq^2}} \left[  \Gamma\Big(\frac{1}{c}, \frac{3c q^2}{2}\Big)
-  \Gamma\Big(\frac{1}{c},   \frac{3c  q^2}{2}   e^{6  c  x}\Big)   \right]  \,
\exp\left(\frac{3}{2}cq^2 -6x\right) \nonumber \\ & & \,\,
\end{eqnarray}

The cosmology of the mixed fluid, at the background level, is determined by its
EoS   parameter  $\omega$   (See  Eq.(\ref{eq:10})).   For  the   models  under
consideration,
\begin{equation}
  \label{eq:29} \omega(x) =  1- \frac{2 c\, r  \,\exp\left[6x -\frac{3}{2}c q^2
(e^{6cx}  -1)\right]}{3c  +  r \sqrt[c]{\frac{2}{3c  q^2}}  e^{3cq^2/2}  \left[
\Gamma\big(\frac{1}{c},    \frac{3cq^2}{2}\big)    -    \Gamma\big(\frac{1}{c},
\frac{3cq^2}{2} e^{6cx} \big)\right]}
\end{equation} Although  the evolution of  EoS parameter $\omega$  depends upon
three parameters,  \emph{i.e.}, $c$, $q$ and  $r$, its value $\omega_0$  at the
present  epoch is  completely given  by the  dimensionless parameter  $r$ which
represents  the  present-epoch  value   of  the  quintessence  self-interaction
potential,
\begin{equation}
  \label{eq:30}
  \omega_0 = 1-\frac{2r}{3}
\end{equation}  Provided $r>2$,  $\omega_0  <-1/3$,  the late-time  accelerated
expansion occurs. In particular, the mixed fluid in ETG of Lyra's geometry will
respectively mimic the  quintessence, cosmological constant and  phantom in the
Einstein's  general relativity  if the  parameter $r$  takes its  value in  the
regions $2<r<3$,  $r=3$ and $r>3$.  However, for  $r\geq 3$, the  EoS parameter
$\omega$ given in Eq.(\ref{eq:29}) increases  monotonically near $x=0$ for real
parameters   $c$  and   $q$,  as   seen  from   the  asymptotic   expansion  of
Eq.(\ref{eq:29}) at small $x$,
\begin{equation}
\label{eq:31} \omega \approx 1-\frac{2r}{3} + \frac{2  r}{3} (9 c^2 q^2 +2r -6)
x
\end{equation}  Such a  $\omega$  conflicts  with our  common  sense about  the
universe evolution. We choose to abandon this possibility. The parameter $r$ is
consequently restricted within the region $2<r<3$ for these models, in turn the
aspired $\omega = -1$ crossing is unavailable.

It is interesting to study the dependence of the evolution of $\omega$ upon the
magnitude of  the dimensionless  coupling parameter  $c$. To  this end  we plot
Eq.(\ref{eq:29})  in FIG.  1.  for  three different  choices  of coupling  $c$,
\emph{i.e.}, $c=2$, $c=1.5$ and $c=1$, and the parameters $q$ and $r$ are fixed
at $q=0.05$  and $r=2.75$, respectively. It  is manifest that the  evolution of
the universe from matter dominant era to the present acceleration phase depends
weakly upon what the  coupling constant $c$ is. However, the  value of $c$ will
strongly influence the  universe evolution in the future. The  larger the value
of $c$ is, the earlier will the universe exit from its accelerated expansion.
\begin{figure}[htb]
\label{fig1}
\begin{center}
\includegraphics[width=3.6in,height=2.7in,angle=0]{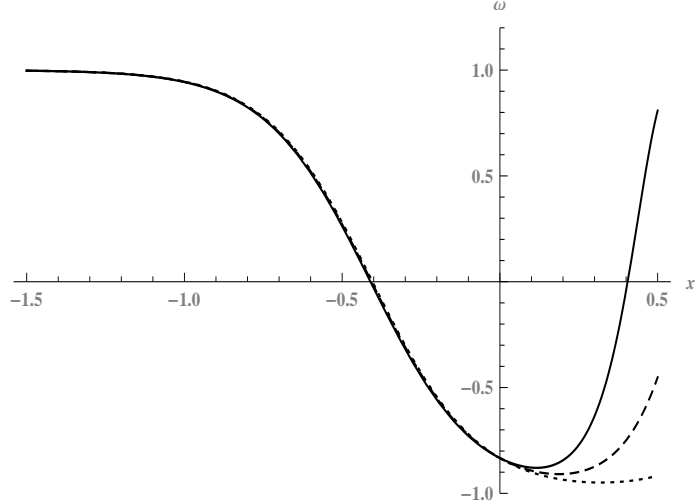} 
\caption[\emph{Evolution of $\omega$ against $x$}]{Evolution of $\omega$ versus
$x$  for  different coupling  parameter  $c$  for interaction  term  ${\mathscr
Q}=3cH\dot{\varphi}^{2}$.  Here we  take  $q=0.05$ and  $r=2.75$ (So  $\omega_0
\approx  -0.83$). The  solid, dashed  and  dotted curves  correspond to  $c=2$,
$c=1.5$ and $c=1$ respectively.}
\end{center}
\end{figure}

The above  is the cosmological implications  of the background dynamics  of the
proposed model in Lyra's geometry. The concordance cosmology is a science based
on precise observations of which lots  are tightly connected to the growth of
perturbations. Thus  we must examine  the stability issue of  the perturbation
modes in  the model under  consideration. According to the  linear perturbation
theory, the stability  of the linear perturbation modes  during their evolution
requires  $c_{s,i}^{2}\geq  0$  for  each  component  fluid  \cite{Garriga1999,
Vikman2005,  Amendola2010}, where  $c_{s,i}^{2} \equiv  \partial p_{i}/\partial
\rho_{i}$ is  its squared  sound speed  at the background  level. In  Lyra's
geometry, the  squared sound  speed  of the  displacement field  is always
definitely positive.   In fact, $c_{s,\theta}^{2}=\omega_{\theta} =  1$. On the
other hand,
\begin{equation}
  \label{eq:32}   c_{s,\varphi}^{2}    =1   +   \frac{2   r    \exp\left[   6x
-\frac{3}{2}cq^{2}(e^{6cx}   -1)\right]}{(c-1)   \left\{    3   +   \frac{r}{c}
\sqrt[c]{\frac{2}{3cq^{2}}}  e^{3cq^{2}/2}  \left[\Gamma  \big  (  \frac{1}{c},
\frac{3}{2}cq^{2}  \big )  - \Gamma\big(\frac{1}{c},  \frac{3}{2}cq^{2} e^{6cx}
\big) \right]\right\}}
\end{equation} The squared sound  speed $c_{s,\varphi}^{2}$ varies continuously
with respect  to e-folding time  unless $c= 1$. For  a positive $c$  ($c\ne 1$),
$c_{s,\varphi}^{2} \rightarrow 1$ when $x\rightarrow \pm \infty$, while,
\begin{equation}
  \label{eq:33}  c_{s,\varphi}^{2}  \approx 1  +  \frac{2r  \left[ 1  +  (6-2r
-9c^{2} q^{2}) x \right]}{3(c-1)}
\end{equation} at  $x\approx 0$. Recall  that $2<r<3$, the squared  sound speed
$c_{s,\varphi}^{2}$  of  the quintessence  scalar  diverges  or takes  negative
values for  $0< c\leq  1$, the corresponding  perturbation modes  are violently
unstable  and  do  not  have   any  physical  significance.  The  stability  of
perturbation  modes is  also  sensitive  to the  initial  velocity  $q$ of  the
quintessence scalar. To ensure a  finite and non-negative $c_{s, \varphi}^{2}$,
the coupling constant $c$ should be restricted  to the region $c>1$, and at the
same time the coefficient of $x$ in  the RHS of Eq.(\ref{eq:33}) should be set
to zero.  Therefore, the  model is  physical acceptable  only if  $c>1$, $q=\pm
\frac{1}{3c}\sqrt{6-2r}$ and $2<r<3$.

\vspace{5mm}
\begin{description}
\item[Case\,2 \,$b\ne 0$ but $c= \tilde{b} = f= 0$\, ]
\end{description} In this case, Eq.(\ref{eq:16}) is translated into,
\begin{equation}
  \label{eq:34}     \left[\frac{d^2\varphi}{dx^2}      -     \frac{b}{\sqrt{6}}
\bigg(\frac{d\varphi}{dx}\bigg)^2 +\sqrt{6} b  \right]H^2 +\left[V_{,\varphi} +
\bigg(\frac{d\varphi}{dx}\bigg) V -\frac{\sqrt{6} b}{3} V \right] =0
\end{equation} Similar to  Case 1, we further assume  that the self-interaction
quintessence potential $V(\varphi)$ satisfies the constraint condition,
\begin{equation}
\label{eq:35} V_{,\varphi} +  \bigg(\frac{d\varphi}{dx}\bigg) V -\frac{\sqrt{6}
b}{3} V = 0
\end{equation} Consequently, the evolution of quintessence scalar in the models
under consideration is also fictitiously independent of the evolution of Hubble
parameter,
\begin{equation}
\label{eq:36}        \frac{d^2\varphi}{dx^2}        -        \frac{b}{\sqrt{6}}
\bigg(\frac{d\varphi}{dx}\bigg)^2 +\sqrt{6} b = 0
\end{equation}  The solution  of Eq.(\ref{eq:36})  which satisfies  the initial
conditions $\varphi(0) =0$ and $d\varphi(0)/{dx} = \sqrt{6} q$ reads,
\begin{equation}
  \label{eq:37}  \varphi(x)  =  -\frac{\sqrt{6}}{b}\ln   \left[  \cosh(b  x)  -
q\sinh(b x) \right]
\end{equation} Plugging Eq.(\ref{eq:37}) into Eq.(\ref{eq:35}) yields,
\begin{equation}
  \label{eq:38}  V(x)  =  \frac{r  H_0^2  }{\left[ \cosh(b  x)  -  q\sinh(b  x)
\right]^2} \exp\left[-6x  + \frac{6(1-q^2)}{b}\,\frac{\sinh(b x)}{\cosh(b  x) -
q\sinh(b x) } \right]
\end{equation} where $r$  is an integration constant which is,  as before, used
to  specify  the  present-epoch  value   of  quintessence  potential,  $V(0)  =
rH_0^2$. Different  from Case  1, for  the current models,  it is  difficult to
express   the    quintessence   potential    as   a   closed    form   function
$V(\varphi)$.  Fortunately,   this  does   not  affect  our   investigation  to
cosmology. With (\ref{eq:38}), we can  easily solve Eq.(\ref{eq:14}) and obtain
the evolution of Hubble parameter in these models,
\begin{equation}
  \label{eq:39}  H^2(x)  =  H_0^2  e^{-6x}\,\left[  1  -\frac{r}{3(1-  q^2)}  +
\frac{r}{3(1- q^2)} \exp\left( \frac{6(1-q^2)}{b}\,\frac{\sinh(b x)}{\cosh(b x)
- q\sinh(b x) } \right) \right]
\end{equation} and then,
\begin{equation}
\label{eq:40} \theta (x) = \frac{4 H_0^2 e^{-6x} (3q^2 + r -3)}{3 \big[ \cosh(b
x) - q\sinh(b x) \big]^2 }
\end{equation}

\vspace{4mm}

Therefore,
\begin{equation}
\label{eq:41}    \omega    =   1    -    \frac{2    (1-q^2)   \,r    \exp\bigg(
\frac{6(1-q^2)}{b}\,\frac{\sinh(b x)}{\cosh(b  x) - q\sinh(b  x) }\bigg)}{\big[
\cosh(b   x)  -   q\sinh(b   x)  \big]^2   \bigg[3(1-q^2)   -r  +r   \exp\left(
\frac{6(1-q^2)}{b}\,\frac{\sinh(b x)}{\cosh(b x) - q\sinh(b x) }\right) \bigg]}
\end{equation} The evolution of the  effective EoS parameter depends upon three
parameters, \emph{i.e.}, $b$,  $q$ and $r$. However, as in  Case 1, its present
value $\omega_0$ is only relevant to $r$,
\begin{equation}
  \label{eq:42} \omega_0 = 1 - \frac{2 r}{3}
\end{equation}  Provided $r>2$,  $\omega_0  <-1/3$,  the late-time  accelerated
expansion occurs.  Notice that the  asymptotic expansion of Eq.(\ref{eq:41}) at
small $x$ is,
\begin{equation}
  \label{eq:43}   \omega   \approx   1-  \frac{2r}{3}   -4   r   x\left[(1-q^2)
-\frac{1}{3}(r -b q)\right]
\end{equation} So  long as  $r< bq+3(1-q^2)$, the  EoS parameter  $\omega$ will
decrease monotonically  near $x=0$  for real parameters  $b$ and  $q$, implying
that the  $w=-1$ crossing is possible.  For a given coupling  constant $b$, the
initial "velocity" $q$ of the quintessence scalar have to take its value in the
region  $(b-\sqrt{b^2  +12})/6  <q  <(b+\sqrt{b^2  +12})/6$  to  guarantee  the
inequality $2<r<bq+3(1-q^2)$. In FIG. 2. we plot the evolution of EoS parameter
$\omega$  in Eq.(\ref{eq:41})  for  three different  choices  of coupling  $b$,
\emph{i.e.}, $b=2.6$, $b=3$  and $b=3.6$, and the parameters $q$  and $r$ are fixed
at  $q=1$ and  $r=2.05$,  respectively. The  remarkable  difference between  the
present case  and Case 1 is  that the EoS in  the present models can  cross the
phantom divide  $\omega =-1$.  Is this quintom scenario reliable?

\begin{figure}[htb]
\label{fig2}
\begin{center}
\includegraphics[width=3.6in,height=2.7in,angle=0]{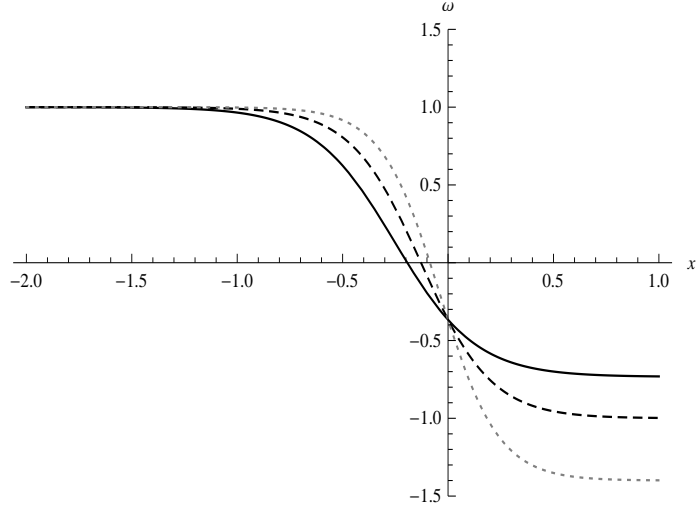} 
\caption[\emph{Evolution of $\omega$ against $x$}]{Evolution of $\omega$ versus
$x$  for  different coupling  parameter  $b$  for interaction  term  ${\mathscr
Q}=3b\theta\dot{\varphi}/{2\sqrt{6}}$.  Here we  take  $q=1$  and $r=2.05$.  The
solid,  dashed  and  dotted  curves   correspond  to  $b=2.6$,  $b=3$  and  $b=3.6$
respectively.}
\end{center}
\end{figure}

It  has been  pointed out  \cite{Vikman2005,  Cai2010a} that  a viable  quintom
scenario can  not be realized  only by virtue  of the parameterization  of EoS.
The stability of the relevant perturbation modes must be ensured also. In other
words, we have to guarantee $c_{s,\varphi}^{2} \geq 0$ at background level. For
simplicity we only consider a special case  $q= 1$. In this case, the solution
to the background dynamics reduces to:
\begin{eqnarray}
\label{eq:44}
&  & \varphi(x) = \varphi(0) + \sqrt{6}x  \\
\label{eq:45}
&  & V(x) = r H_{0}^{2} e^{2(b-3)x} \\
\label{eq:46}
&  & H^{2} = H_{0}^{2}e^{-6x} \left[1 + \frac{r}{b}(e^{2bx} -1) \right] \\
\label{eq:47}
&  & \theta(x) = \frac{4}{3}rH_{0}^{2}e^{2(b-3)x} \\
\label{eq:48}
&  & \omega = 1- \frac{2br e^{2bx}}{3\big[b + r (e^{2bx} -1) \big]}
\end{eqnarray}
The energy density and pressure of the quintessence scalar read,
\begin{equation}
  \label{eq:49}
\rho_{\varphi} = \frac{H_{0}^{2}}{b}e^{-6x} \big[ 3(b-r) + (3+ b) r
e^{2bx} \big] \,,\,\,\,
p_{\varphi} =  \frac{H_{0}^{2}}{b}e^{-6x} \big[ 3(b-r) + (3- b) r
e^{2bx} \big]
\end{equation}
Hence,
\begin{equation}
\label{eq:50}   c_{s,\varphi}^{2}   =   \frac{\partial   p_{\varphi}}{\partial
\rho_{\varphi}} = \frac{(3-b)^{2}  r e^{2bx} + 9(b-r)}{(9-b^{2}) r  e^{2bx} + 9
(b-r)}
\end{equation} The  physical acceptance requires  $\omega <-1/3$ for  $x\geq 0$
and $c_{s,\varphi}^{2} \geq 0$. Obviously,  both inequality can be satisfied if
$2<r <b \leq  3$. With respect to such a  parameter constraint, however, $1\geq
\omega \geq  1-2b/3$, the  quintom scenario  where the EoS  of mixed  fluid can
cross the cosmological constant boundary is  forbidden. It seems that the no-go
theorem \cite{Vikman2005, Hu2005, Caldwell2005, Zhao2005, Kunz2006, Xia2008} is
valid  also  for  a generic  $q$  so  that  a  reasonable quintom  scenario  is
unavailable in the present model.

\begin{figure}[htb]
\label{fig3}
\begin{center}
\includegraphics[width=3.6in,height=2.7in,angle=0]{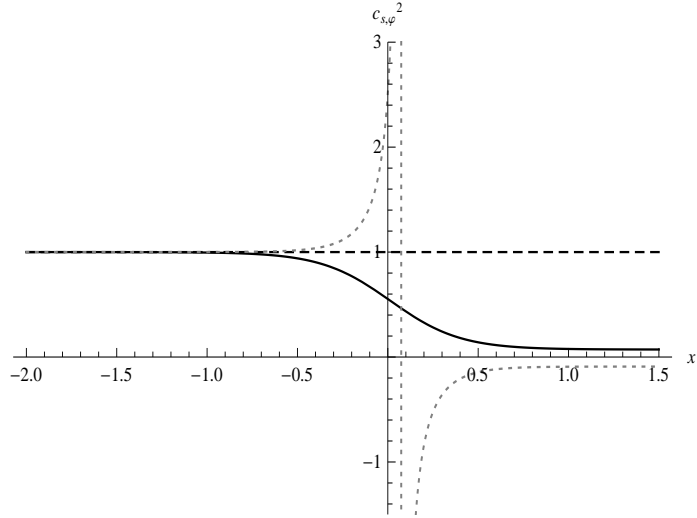} 
\caption[\emph{Evolution    of    $\omega$     against    $x$}]{Evolution    of
$c_{s,\varphi}^{2}$  versus  $x$  for  different  coupling  parameter  $b$  for
interaction  term  ${\mathscr Q}=3b\theta\dot{\varphi}/{2\sqrt{6}}$.   Here  we
take $q=1$  and $r=2.05$.  The  solid, dashed  and dotted curves  correspond to
$b=2.6$,   $b=3$  and   $b=3.6$   respectively.  In   the   first  two   cases,
$c_{s,\varphi}^{2}$   is    finite   and   positive.   In    the   last   case,
$c_{s,\varphi}^{2}$ diverges during its  evolution, implying the instability of
the relevant perturbation modes. }
\end{center}
\end{figure}

\vspace{3mm}
\begin{description}
\item[Case\,3 \,$\tilde{b}\ne 0$ but $c= b = f= 0$\, ]
\end{description} In this case the  interaction between the quintessence scalar
and  displacement  field  is ${\mathscr  Q}=  \frac{9}{2}\tilde{b}H\theta$.  We
choose Eqs.(\ref{eq:14}),  (\ref{eq:16}) and (\ref{eq:17})  to form the  set of
independent equations. The latter two can be recast as,
\begin{eqnarray}
\label{eq:51} &  & H^2\left[\frac{d^2\varphi}{dx^2} + \bigg(3  + \frac{1}{2H^2}
\frac{dH^2}{dx}    \bigg)\frac{d\varphi}{dx}     \right]    +    \frac{dV}{dx}-
\frac{9}{2}\tilde{b}    \theta   =    0   \\    &   &    \frac{d\theta}{dx}   +
6(1-\tilde{b})\theta = 0
\label{eq:52}
\end{eqnarray} among which,  Eq.(\ref{eq:52}) is easily to  solve. By assigning
the  initial  condition  $\theta  (0)=   4(\tilde{b}-1)  sH_0^2$,  with  $s$  a
dimensionless constant, we have the solution of Eq.(\ref{eq:52}) as follow,
\begin{equation}
  \label{eq:53} \theta(x) = 4(\tilde{b} -1)s \,H_0^2 e^{6(\tilde{b}-1)x}
\end{equation} We  further assume  that in the  models under  consideration the
self-interaction potential of the quintessence scalar possesses property,
\begin{equation}
  \label{eq:54} \frac{dV}{dx}- \frac{9}{2}\tilde{b} \theta = 0
\end{equation} Under this assumption, Eq.(\ref{eq:51}) is reduced to,
\begin{equation}
  \label{eq:55}    \frac{d^2\varphi}{dx^2}   +    \bigg(3   +    \frac{1}{2H^2}
\frac{dH^2}{dx} \bigg)\frac{d\varphi}{dx} = 0
\end{equation} Substitution of Eq.(\ref{eq:53}) into (\ref{eq:54}) yields,
\begin{equation}
  \label{eq:56} V(x)  = H_0^2  \left[r +  3\tilde{b} s  (e^{6(\tilde{b}-1)x} -1
)\right]
\end{equation} where $r$ is an integration  constant. As in the previous cases,
this  parameter  characterizes  the  present-epoch value  of  the  quintessence
potential. With  Eq.(\ref{eq:56}), we  can obtain the  evolution of  the Hubble
parameter by solving Eq.(\ref{eq:14}). The result is,
\begin{equation}
  \label{eq:57}  H^2   =  H_0^2   \left[  (1  +   \tilde{b}s  -s)   e^{-6x}  +s
e^{6(\tilde{b}-1)x} -\tilde{b}s \right] + \frac{r}{3}H_0^2 (1- e^{-6x})
\end{equation} Plugging Eq.(\ref{eq:57}) into (\ref{eq:55}) gives,
\begin{equation}
  \label{eq:58}  \frac{d\varphi}{dx}  =  \frac{q}{\sqrt{3  +  (r  -3\tilde{b}s)
(e^{6x}-1) + 3s (e^{6\tilde{b}x}-1) }}
\end{equation} where  the integration  constant $q$ is  related to  the initial
"velocity"  of the  quintessence scalar  by $d\varphi(0)/{dx}=q/\sqrt{3}$.  The
cosmology of the considered models at  the background level is described by the
following effective EoS parameter:
\begin{equation}
  \label{eq:59}   \omega   =   1-  2e^{6x}\left[\frac{r   +   3\tilde{b}s   (1-
e^{6(\tilde{b}-1)   x})}{3    +   (3\tilde{b}s    -r)   (1-    e^{6x})   -3s(1-
e^{6\tilde{b}x})} \right]
\end{equation} The  EoS parameter seems  not to depend  upon the choice  of the
initial "velocity" $q$  of the quintessence scalar, but upon  the initial value
$s$ of  the displacement  field instead.  This is,  however, merely  an optical
illusion. The consistence of the above results with Eq.(\ref{eq:15}) requires,
\begin{equation}
\label{eq:60} (\tilde{b}-1) s = \frac{q^2 +6(r-3)}{18}
\end{equation} The present-epoch  value of EoS parameter is still  given by the
same formula  as either Eq.(\ref{eq:31})  or (\ref{eq:42}), which  depends only
upon the parameter $r$,
\begin{equation}
  \label{eq:61} \omega_0 = 1 - \frac{2r}{3}
\end{equation} Provided  $r>2$, the universe  is destined to enter  a late-time
acceleration phase.

To examine the  stability of the relevant perturbation modes,  we calculate the
squared sound speed of the quintessence scalar. The result is,
\begin{equation}
\label{eq:62}  c_{s,\varphi}^{2} =  \frac{3-r  +  3(\tilde{b}-1)s +  3\tilde{b}
(\tilde{b}-1)s     e^{6\tilde{b}x}}{3-r    +     3(\tilde{b}-1)s    -3\tilde{b}
(\tilde{b}-1)s e^{6\tilde{b}x}}
\end{equation} As  long as  $(\tilde{b}-1)s \ne 0$,  $c_{s,\varphi}^{2}$ either
diverges or becomes negative during its  evolution, which will give rise to the
unstable and then the unacceptable perturbation modes.  On the other hand, when
$(\tilde{b}-1)s  =  0$,   which  occurs  for  either   $\tilde{b}=1$  or  $s=0$,
$c_{s,\varphi}^{2}=1$,  the relevant  perturbation modes  might evolve  stably.
For $\tilde{b}=1$, the solution to the background dynamics reduces to:
\begin{eqnarray}
\label{eq:63} & & \varphi(x)  = -\sqrt{\frac{2}{3}} \tanh^{-1}\sqrt{\frac{3-r +
r e^{6x}}{3-r}} \\
\label{eq:64} & & V(x) = r H_{0}^{2} \\
\label{eq:65} & & H^{2} = \frac{1}{3} H_{0}^{2} \Big[r + (3-r)e^{-6x} \Big] \\
\label{eq:66} & & \theta(x) = 0
\end{eqnarray} and in particular,
\begin{equation}
\label{eq:67} \omega = -1 + \frac{2(3-r)}{3-r + r e^{6x}}
\end{equation} In this case, the  displacement field is effectively absent, and
the self-interaction potential of the quintessence scalar plays the role of the
cosmological constant.  The  potential parameter $r$ must be  restricted to the
region  $2<r \leq  3$, otherwise  $\omega$ will  diverge during  its evolution.
Consequently, crossing the phantom divide $\omega  =-1$ in the present model is
practically  impossible.  Fig.  4.  shows  the evolution  of  EoS parameter  in
Eq.(\ref{eq:67})   for  three   choices  of   the  potential   parameter  $r$,
\emph{i.e.,} $r =2.6$, $r =2.9$ and  $r=3$, with the coupling constant fixed at
$\tilde{b}=1$.  It appears that the  accelerated expansion in these models will
last for a very  long time. Besides, the larger the parameter  $r$ is, the more
closely the quintessence scalar resembles the cosmological constant.

\begin{figure}[htb]
\label{fig4}
\begin{center}
\includegraphics[width=3.6in,height=2.7in,angle=0]{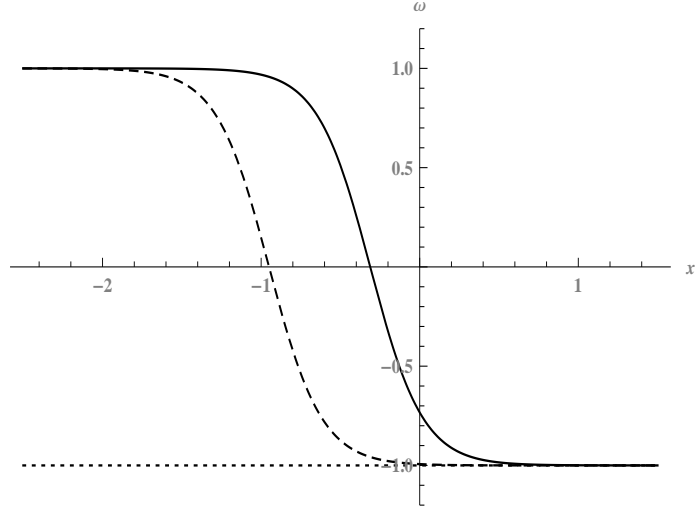} 
\caption[\emph{Evolution of $\omega$ against $x$}]{Evolution of $\omega$ versus
$x$ for  different potential  parameter $r$ if  the interaction  term is
${\mathscr Q}=\frac{9}{2}\tilde{b}  H\theta$ with $\tilde{b}=1$. The solid,
dashed and dotted  curves correspond to $r=2.6$, $r=2.9$ and $r=3$, respectively.}
\end{center}
\end{figure}

\vspace{5mm}
\begin{description}
\item[Case\,4 \,$f\ne 0$ but $b= \tilde{b} = c= 0$\, ]
\end{description} In this case, the interaction between the quintessence scalar
and the  displacement field is assumed  to be proportional to  the quintessence
self-interaction  potential, ${\mathscr  Q}=6 f  H V(\varphi)$.  Under such  an
assumption, Eqs.(\ref{eq:16}) and (\ref{eq:17}) become,
\begin{eqnarray}
\label{eq:68} &  & \frac{d\varphi}{dx}\,  \left[ H^2  \frac{d^2\varphi}{dx^2} +
\bigg(3H^2 +  \frac{1}{2}\frac{dH^2}{dx} \bigg)  \frac{d\varphi}{dx}\right]\, +
\frac{dV}{dx} -6f V= 0 \\
\label{eq:69} & & \frac{d\theta}{dx} + 6\theta -8f V = 0
\end{eqnarray} As before, we further  assume that the quintessence potential is
defined by condition,
\begin{equation}
  \label{eq:70} \frac{dV}{dx} -6f V = 0
\end{equation} This implies that the quintessence self-interaction potential in
the considered models is of the form,
\begin{equation}
  \label{eq:71} V(x) = r H_0^2 e^{6f x}
\end{equation}  where  a  real  parameter  $r$  is  used  to  characterize  the
present-epoch value of the potential and  $H_0$ stands for the present value of
the Hubble  parameter. Substitution  of Eq.(\ref{eq:71})  into Eq.(\ref{eq:14})
yields,
\begin{equation}
  \label{eq:72} H^2 =  H_0^2 e^{-6x}\left[1+ \frac{r}{3(f+1)} \left(e^{6(f+1)x}
-1 \right)\right]
\end{equation} Furthermore,  we can  obtain the  evolution of  the quintessence
scalar by plugging Eqs.(\ref{eq:71}) and (\ref{eq:72}) into Eq.(\ref{eq:68}),
\begin{equation}
  \label{eq:73}                           \varphi(x)                          =
\frac{q}{\sqrt{3(f+1)(3+3f-r)}}\left[\tanh^{-1}\sqrt{\frac{3(f+1)}{3+3f-r}}   -
\tanh^{-1}\sqrt{\frac{3(f+1) +r (e^{6(f+1)x} -1)}{3+3f-r}}\right]
\end{equation} where the parameter $q$ is the integration constant which can be
interpreted   as   the   initial   velocity   of   the   quintessence   scalar,
$d\varphi(0)/{dx}  =q$. Finally,  the evolution  of the  displacement field  is
obtained    from    Eqs.(\ref{eq:15}),   (\ref{eq:71}),    (\ref{eq:72})    and
(\ref{eq:73}),
\begin{equation}
  \label{eq:74} \theta  (x) =  \frac{2}{3} H_0^2  e^{-6x}\,\left[q^2 -6  +2r \,
\frac{fe^{6(f+1)x}+1}{f+1} \right]
\end{equation}

The effective  EoS parameter  of the quintessence  scalar and  the displacement
field in the models under consideration reads,
\begin{equation}
  \label{eq:75}  \omega  =  1-  \frac{2(f+1)   r  \,  e^{6(f+1)x}}{3(f+1)  +  r
(e^{6(f+1)x} -1)}
\end{equation} The present-epoch EoS parameter  takes the same formula as those
in the previous three cases,
\begin{equation}
  \label{eq:76} \omega_0 = 1- \frac{2r}{3}
\end{equation} So the late-time cosmological acceleration is available in these
models if  $r>2$.  Different from the  EoS parameter in the  present case which
depends only  upon the coupling constant  $f$ and the potential  parameter $r$,
the  squared sound  speed  of the  quintessence scalar  depends  also upon  the
initial velocity $q$ of the quintessence scalar,
\begin{equation}
  \label{eq:77}
  c_{s,\varphi}^{2} = \frac{q^{2} + 2fr e^{6(f+1)x}}{q^{2} -2fr e^{6(f+1)x}}
\end{equation} Stability  condition $c_{s,\varphi}^{2}  \geq 0$  requires $f=0$
and $q\ne 0$. Therefore, the model is physically acceptable only if there is no
interaction between  the quintessence scalar  and the displacement  field. When
$f=0$, $c_{s,\varphi}^{2} =1$, the solution  to the background dynamics reduces
to:
\begin{eqnarray}
\label{eq:78}
&   & \frac{d\varphi}{dx} = \frac{\sqrt{3} q}{\sqrt{3-r + r e^{6x}}} \\
\label{eq:79}
&   & V(x) = r H_{0}^{2} \\
\label{eq:80}
&   & H^{2} = \frac{1}{3}H_{0}^{2} \Big[r + (3-r)e^{-6x} \Big]  \\
\label{eq:81}
&   & \theta(x) = \frac{2}{3}H_{0}^{2} e^{-6x} (q^{2} -2r -6)
\end{eqnarray} Because  of $\theta (x) \geq  0$, the value of  parameter $q$ is
restricted to  either $q  \geq \sqrt{2(3-r)}$ or  $q\leq -  \sqrt{2(3-r)}$. For
$f=0$, Eq.(\ref{eq:75}) reduces to:
\begin{equation}
\label{eq:82} \omega = 1- \frac{2r e^{6x}}{3-r + r e^{6x}}
\end{equation} To  have $\omega <-1/3$ at  $x=0$, $r>2$. To avoid  the possible
divergence of $\omega$ during its evolution, $r \leq 3$.  FIG.  5.  illustrates
the dependence  of the  EoS parameter in  Eq.(\ref{eq:82}) upon  the potential
parameter $r$. Obviously, $-1 \leq \omega  \leq 1$, crossing the phantom divide
$\omega  =-1$ is  prohibited  once more  in the  quintessence  model in  Lyra's
geometry.

\begin{figure}[htb]
\label{fig5}
\begin{center}
\includegraphics[width=3.6in,height=2.7in,angle=0]{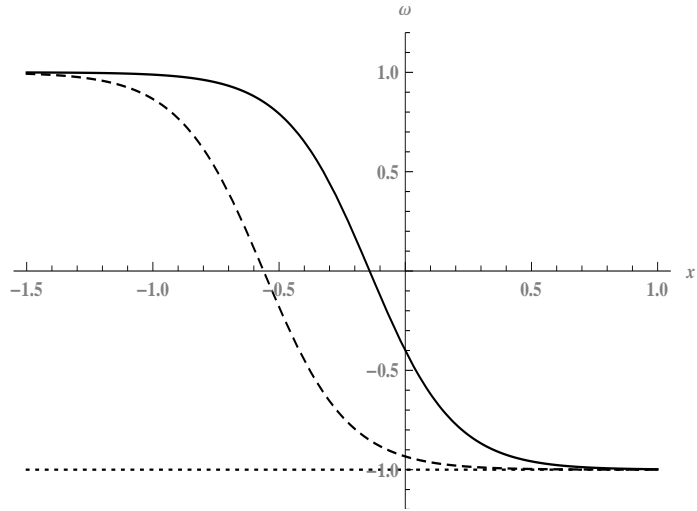} 
\caption[\emph{Evolution of $\omega$ against $x$}]{Evolution of $\omega$ versus
$x$ for different potential parameter $r$  if there is no interaction between
the quintessence scalar and the displacement field. The solid, dashed and
dotted  curves correspond to $r=2.1$, $r=2.9$ and $r=3$, respectively.}
\end{center}
\end{figure}

\section{Conclusion}

In  this work,  we have  established  \emph{four} classes  of the  quintessence
models in ETG  in Lyra's geometry.  The classification of  these models depends
upon  how the  quintessence  scalar $\varphi$  interacts  with the  geometrical
displacement  field  $\theta$  (or  $\beta_\mu$). The  mixture  of  interacting
quintessence  scalar and  the  displacement field  supplies  as a  cosmological
perfect fluid which can cause late  time accelerated expansion of our universe.
Owing to the subtle choices of the quintessence self-interaction potential, all
quantities relevant to the study  of cosmology, including the Hubble parameter,
the displacement field, the time derivatives of the quintessence scalar and the
potential  itself, are  expressed as  closed  form functions  of efolding  time
$x=\ln(a)$, and so is the effective  EoS parameter $\omega$ of the mixed fluid.
The evolution of $\omega$ is different for the quintessence models of different
classes, which depends  also upon the present values of  the time derivative of
quintessence scalar, its potential and  what the coupling constant is. However,
today's  $\omega$  does  only  depend  upon  the  present-epoch  value  of  the
quintessence  potential. The  appearance of  the displacement  field in  Lyra's
geometry    improves     greatly    the    late-time    evolution     of    the
quintessence-displacement  mixture,   however,  crossing  the   phantom  divide
$\omega=-1$  is still  forbidden in  these  models by  the necessary  condition
$c_{s,  \varphi}^2  \geq  0$  to  ensure  the  stability  of  the  cosmological
perturbations. Establishing a reliable quintom  scenario remains a challenge in
ETG in Lrya's geometry .

In the proposed quintessence models in  ETG in Lyra's geometry, we have defined
the quintessence scalar by making some careful choices for its self-interaction
potential  and  interaction terms  with  the  geometrical displacement.   These
choices, in this  paper, are mainly motivated by the  requirement to obtain the
analytical solutions of the modified  Friedmann equations. More important issue
that remains unsolved  is to investigate the physics behind  these choices. Why
the EoS parameter of the quintessence scalar at present epoch depends only upon
the parameter $r$ for all four kinds  of models is also a mystery. In addition,
the characteristic behaviour that the effective EoS parameter $\omega$ decrease
from $\omega  \approx 1$  in the  past conflicts  with the  well-known Big-Bang
diagram. In our $\omega\sim x$ figures the dust-dominant phase does not form an
expected  plateau. To  be more  realistic,  the pressureless  cold dark  matter
components have to be introduced into the model buildings also.

\section{Acknowledgements}

We are grateful to the discussions with Yi-Fu Cai, Jian-Xin Lu, Wenzhong Liu and
Jun Ouyang. We also thank Alexander Vikman for his comments on no-go theorem
crossing the cosmological constant boundary.

%

\end{document}